\documentclass[reprint, aip, cha, amssymb, amsmath, amsfonts, a4paper, floatfix, superscriptaddress, showpacs]{revtex4-1}

\usepackage{rotating, xspace, units}

\newcommand{\kl}[1]{\left( #1 \right)}

\newcommand{\kle}[1]{\left[ #1 \right]}

\newcommand{\defi}{\mathrel{\mathop:}=}

\newcommand{\Epi}{\affiliation{Department of Epileptology, University of Bonn, Sigmund-Freud-Stra{\ss}e~25, 53105~Bonn, Germany}}
\newcommand{\HISKP}{\affiliation{Helmholtz Institute for Radiation and Nuclear Physics, University of Bonn, Nussallee~14--16, 53115~Bonn, Germany}}
\newcommand{\IZKS}{\affiliation {Interdisciplinary Center for Complex Systems, University of Bonn, Br\"uhler Stra\ss{}e~7, 53175~Bonn, Germany}}
\newcommand{\ICBM}{\affiliation {Theoretical Physics/Complex Systems, ICBM, Carl von Ossietzky University of Oldenburg, \\Carl-von-Ossietzky-Stra\ss{}e~9--11, Box~2503, 26111~Oldenburg, Germany}}
\newcommand{\RNS}{\affiliation{Research Center Neurosensory Science, Carl von Ossietzky University of Oldenburg,\\ Carl-von-Ossietzky-Stra\ss{}e~9--11, 26111~Oldenburg, Germany}}

\begin{document}

\title{Complexity and irreducibility of dynamics on networks of networks} 

\author{Leonardo Rydin Gorj\~ao}
\email{\texttt{leonardo.rydin@gmail.com}}
\Epi \HISKP

\author{Arindam Saha}
\email{\texttt{arindam.saha@uol.de}}
\ICBM

\author{Gerrit Ansmann}
\email{\texttt{gansmann@uni-bonn.de}}
\Epi \HISKP \IZKS

\author{Ulrike Feudel}
\email{\texttt{ulrike.feudel@uni-oldenburg.de}}
\ICBM \RNS

\author{Klaus Lehnertz}
\email{\texttt{klaus.lehnertz@ukbonn.de}}
\Epi \HISKP \IZKS

\begin{abstract}
We study numerically the dynamics of a network of all-to-all-coupled, identical sub-networks consisting of diffusively coupled, non-identical FitzHugh--Nagumo oscillators.
For a large range of within- and between-network couplings, the network exhibits a variety of dynamical behaviors, previously described for single, uncoupled networks.
We identify a region in parameter space in which the interplay of within- and between-network couplings allows for a richer dynamical behavior than can be observed for a single sub-network.
Adjoining this atypical region, our network of networks exhibits transitions to multistability.
We elucidate bifurcations governing the transitions between the various dynamics when crossing this region and discuss how varying the couplings affects the effective structure of our network of networks.
Our findings indicate that reducing a network of networks to a single (but bigger) network might be not accurate enough to properly understand the complexity of its dynamics.
\end{abstract}

\maketitle
\begin{quotation}
Many natural systems ranging from ecology to the neurosciences can be described as networks of networks.
An example are interacting patches of neural tissue, where each patch constitutes a sub-network.
In such a configuration and other models, the sub-networks are often assumed to be identical, but consisting of non-identical units.
A question arising for such networks of networks is as to what extent their dynamics can be reduced to a more simple network by aggregating parts of the network.
We here explore this question by investigating numerically the dynamics of a network of all-to-all coupled, identical networks consisting of diffusively coupled, non-identical excitable FitzHugh--Nagumo oscillators.
Intriguingly, we identify a small region of the parameter space spanned by the within- and between-network coupling strength that allows for a richer dynamical behavior than what can be observed for a single sub-network.
\end{quotation}

\frenchspacing

\section{Introduction}
Over the past two decades, complex networks have proven valuable to improve our understanding of organizational principles and dynamics of spatially extended systems, which led to broad application in diverse scientific fields \cite{boccaletti2006, arenas2008, donges2009, barthelemy2011, barabasi2011, newman2012, baronchelli2013, lehnertz2014, stam2014, pastorsatorras2015, bassettsporns2017}.
Taking into account that many real systems of different or similar nature usually interact with each other, the network approach was recently extended leading to novel concepts such as interdependent networks, interconnected networks, networks of networks, multi-layered networks, or multiplex networks \cite{kurant2006, gao2012, agostini-scala2014, boccaletti2014, gao2014, kivelae2014, bartsch2015, kenett2015, maluck2015}.
Studies on networks of networks have revealed novel characteristics that could not be observed for single networks. 
This applies to structural properties where the connectivity in such network of networks is the most important determinant as well as to dynamical properties, where particularly a synchronized dynamics of several or all nodes attracts most interest.
Networks of networks have been proven to be robust with respect to perturbations, notably to cascading failures \cite{parshani2010, gao2011, gao2012, majdandzic2016, radicchi2017} and even if consisting of two functionally identical coupled networks~\cite{wang2016}.

Considering the dynamics on networks of networks, novel synchronization phenomena could be identified such as breathing synchronization \cite{louzada2013}, explosive synchronization \cite{boccaletti2014, zhang2015}, intra-layer \cite{gambuzza2015} and inter-layer \cite{sevilla2016, leyva2017} synchronization, asymmetry-induced \cite{zhang2017} and cluster \cite{singh2017} synchronization.
Furthermore, it has been shown that a variety of complex dynamics like traveling waves, multistability, quasi-periodicity, and chaos can emerge \cite{luke2014, sonnenschein2015}. 
The concept of the master stability function has been extended to multi-layer networks \cite{delgenio2016}. 

When investigating networks of networks the question arises as to what extent the dynamics of such a complex network can be reduced by aggregating parts of the network.
This question of reducibility arises particularly in the case when the network of networks is constructed by identical sub-networks.
For multi-layered networks, an algorithm based on the von Neumann entropy has been proposed to reduce the number of layers \cite{dedomenico2015,dedomenico2015b}.
However, these studies focus on structural properties of the network and the question of reducibility has rarely been discussed from the point of view of the dynamics on such networks \cite{pikovsky2008,diakonova2016,pietras2016}.
Here, we follow the latter studies' line of research and investigate the dynamics of a network of all-to-all coupled, identical networks consisting of diffusively coupled, non-identical excitable units.
Such a configuration is often used in neuroscientific studies of groups of interacting neuronal networks \cite{montbrio2004, barreto2008, kiss2008, olmi2010} and can be regarded as the simplest representation of a network of networks.
We find that our network of networks not only allows for a richer dynamical behavior than the corresponding single networks but also provides important clues about the reducibility of networks of networks.

\begin{figure*}
\includegraphics[width=\linewidth]{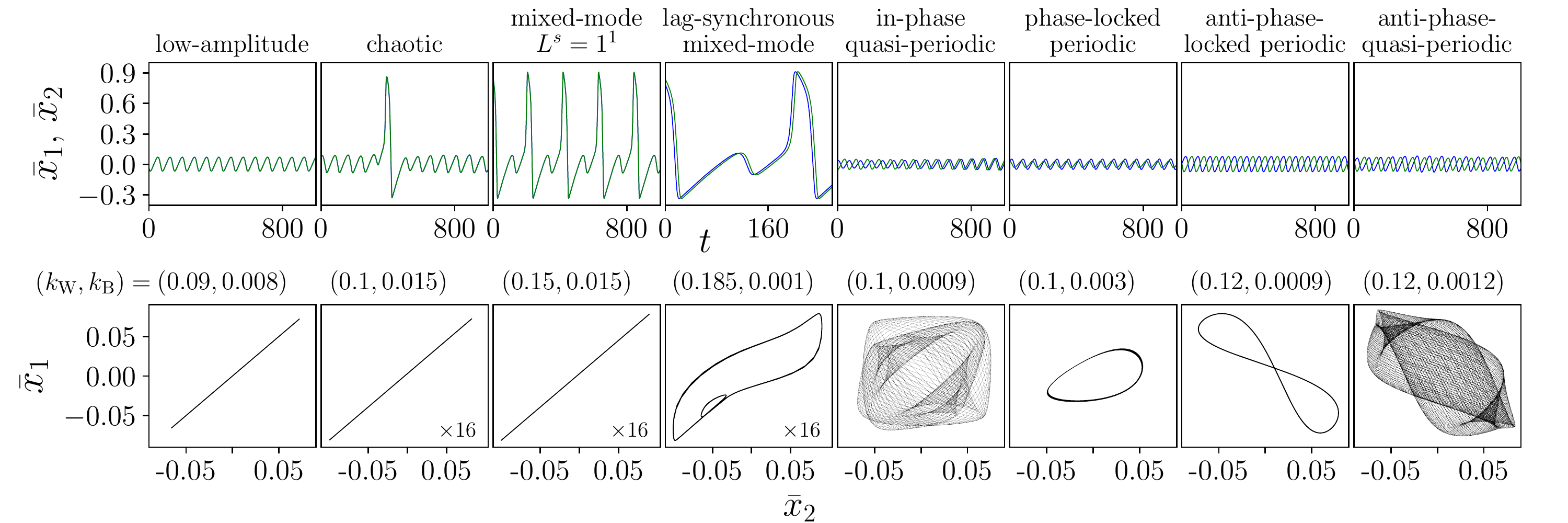}
	\caption{Exemplary temporal evolutions of the average value of the first dynamical variables \(\bar{x}_1\) and \(\bar{x}_2\) (top row) and projections of their trajectories in phase space from \(10^4\) time units (bottom row) for eight choices of coupling strengths \((k_\text{W},k_\text{B})\).
	Sub-networks consisted of $n=10$ units each.
	The headers indicate the designation of each dynamics or oscillation.
	``\(\times 16\)'' indicates that the shown phase-space projection has to be scaled by this factor.
	}
	\label{fig:plots}
\end{figure*}

\section{The Model}\label{sec:model}
We consider a network (or system) that consists of two identical sub-networks, each of which consists of \(n\)~diffusively coupled FitzHugh--Nagumo oscillators~\cite{fitzhugh1961,nagumo1962} (or units).
The dynamics of unit~\(i\) in sub-network~\(q\) is governed by the following differential equations:
	\begin{equation}
		\begin{alignedat}{1}\label{Eq:FHN}
			\dot{x}_{qi} & = x_{qi} (a-x_{qi}) (x_{qi}-1) - y_{qi} + K_\text{W} + K_\text{B},\\
			\dot{y}_{qi} & = b_i x_{qi} - c y_{qi},
		\end{alignedat}
	\end{equation}
where \(x_{qi}\) is the excitatory and \(y_{qi}\) the inhibitory variable, \(a\), \(b\), and~\(c\) are control parameters and  \(K_\text{W}\) and \(K_\text{B}\) are the coupling terms.

Following previous studies~\cite{ansmann2013,ansmann2016}, we fix \(a=-0.0276\) and \(c=0.02\) for all units and choose
	\begin{equation}\label{Eq:b}
		b_i \defi 0.006 + \frac{i-1}{n-1}\times 0.008,
	\end{equation}
i.e., inhomogeneously.
Within a sub-network, the units are coupled completely and diffusively:
	\begin{equation}\label{Eq:internal_coupling}
		K_\text{W} \defi \frac{k_\text{W}}{n-1} \sum\limits_{j=1}^{n} (x_{qj} - x_{qi}),
	\end{equation}
where \(k_\text{W}\) is the \textit{within-network coupling strength.}
Note that with the aforementioned setting of control parameters each unit would exhibit high-amplitude oscillations if units were uncoupled. 
The coupling suppresses these oscillations so that each sub-network acts like an excitable system.

The sub-networks are structural copies of each other.
They are coupled to each other diffusively via their mean fields:
	\begin{equation}\label{Eq:external_coupling}
		K_\text{B} \defi \frac{k_\text{B}}{n} \sum\limits_{j=1}^{n} (x_{rj} - x_{qi}),
	\end{equation}
where \(r=1\) if \(q=2\) and \(r=2\) if \(q=1\), and \(k_\text{B}\) is the \textit{between-network coupling strength.}

The system entails enough complexity to give rise to several different dynamics, made explicit in Sec.~\ref{partIII}.
The study of a single sub-network is at the same time comprehensible enough to distinguish the existing dynamics and transitions of regimes, extensively studied previously~\cite{ansmann2013,ansmann2016}.

The dynamics (Eq.~\ref{Eq:FHN}) was numerically integrated using an adaptive fifth-order Runge--Kutta--Dormand--Prince procedure~\cite{ansmann2017Jit}.
For each of the following observations and analyses, at least \(10^4\) initial time units were discarded.
The choice of the initial conditions (near the attractor) had no influence on our observations.
In Sec.~\ref{sec:Multistability}, different random initial conditions were used per realization.

We choose as the main observable the mean of the excitatory variables for each sub-network:
\(\bar{x}_q\kl{t} \defi \tfrac{1}{n} \sum_{i=1}^n x_{qi}\kl{t}\).
These capture the oscillatory behavior of each sub-network.
We also employ the mean field of the entire network: \(\bar{x} \defi \kl{\bar{x}_1 + \bar{x}_2}/2\).

\begin{figure}
\includegraphics[width=\linewidth]{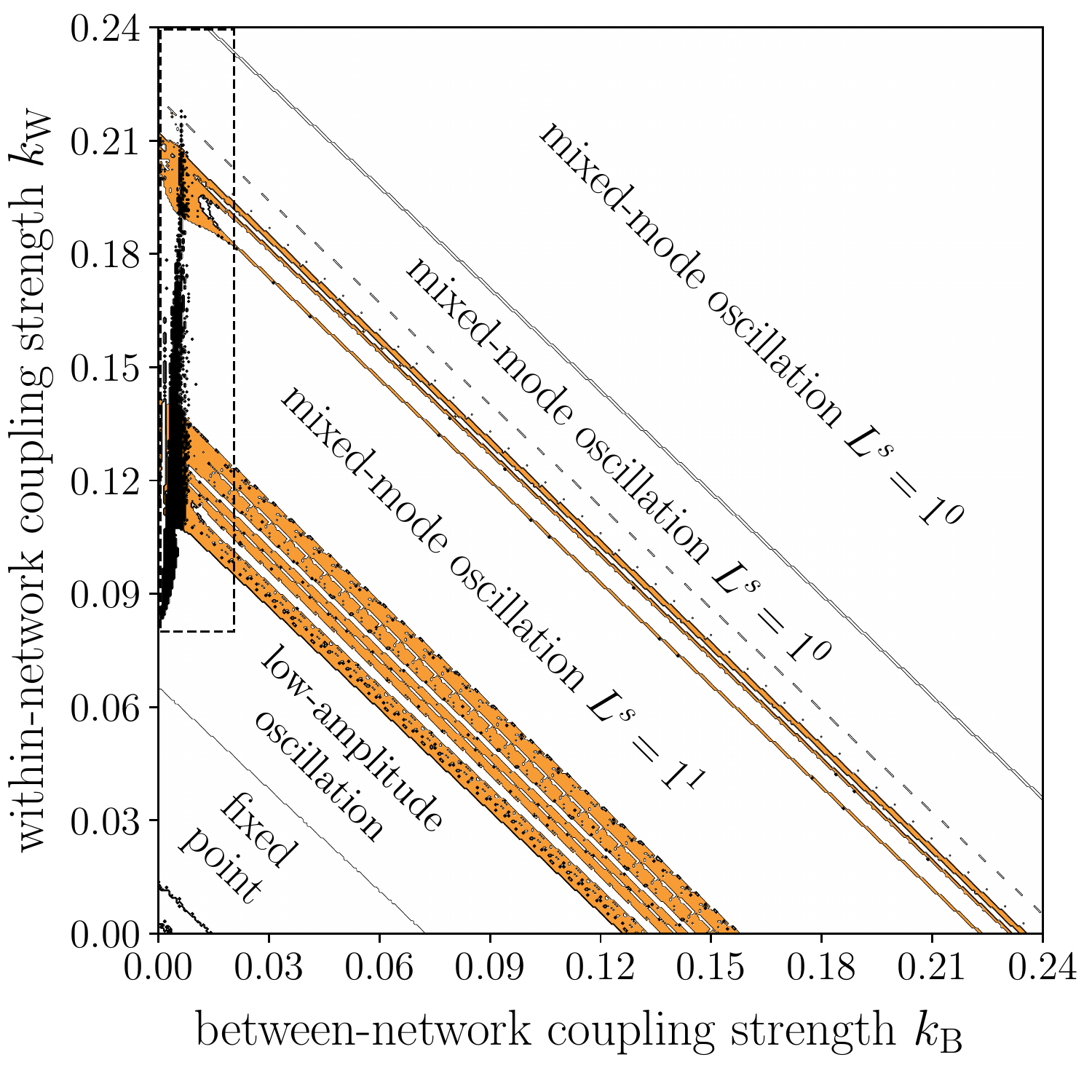}
	\caption{Dynamical regimes in parameter space \((k_\text{B},k_\text{W})\).
	The dynamics observed during \(5\times10^4\) time units are indicated by text or color.
	Orange indicates chaotic dynamics.
	The black region --~the atypical region~-- is composed of several different dynamics; Fig.~\ref{fig:blob2} shows a zoom into the region marked with a dashed box.
	Due to the existence of multistability random initial conditions are chosen at each pair \((k_\text{W},k_\text{B})\).}
	\label{fig:blob1}
\end{figure}

\section{Overview of Dynamical Regimes}\label{partIII}
We first consider the case \(n=10\), i.e., two sub-networks of ten units each.
This choice permits enough inhomogeneity of the sub-networks --~arising from the different values of~\(b_i\)~-- whilst keeping the system coupling structure simple.
We investigate the dynamics in the parameter space spanned by the two coupling strengths \(k_\text{W}\) and~\(k_\text{B}\).

In Fig.~\ref{fig:plots}, we show --~for each sub-network~\(q\)~-- typical time series of the average value of the first dynamical variables, \(\bar{x}_q\).
The sub-networks show a variety of complex dynamical behaviors~\cite{karnatak2014}, including low-amplitude oscillations, chaotic dynamics, different types of mixed-mode oscillations, in- and anti-phase quasi-periodic, as well as phase-locked and anti-phase-locked dynamics.

For \(0.0015 \lesssim (k_\text{W}+k_\text{B}) \lesssim 0.064\), the trivial fixed point at \((\bar{x}_1,\bar{y}_1,\bar{x}_2,\bar{y}_2)=(0,0,0,0)\) is stable and all solutions converge to this point.
For \(0.064 \lesssim (k_\text{W}+k_\text{B}) \lesssim  0.123\), regular low-amplitude oscillations emerge from the fixed point.
Chaotic dynamics are identified by a positive Lyapunov exponent and are also characterized by the irregularity of the high-amplitude peaks observed.
In between and beyond the chaotic regimes (cf. Fig.~\ref{fig:blob1}), we observe mixed-mode oscillations, i.e., periodic dynamics featuring high- and low-amplitude oscillations.
The types of mixed-mode oscillations are specified via the short-hand designation \(L^s\), where \(L\)~denotes the number of high-amplitude and \(s\)~the number of low-amplitude oscillations during one period.
In a comparative sense, low-amplitude dynamics are a special case of mixed-mode oscillations corresponding to the case~\(L^s=0^1\).
Likewise, the perpetual high-amplitude oscillations observed at a high-coupling strength correspond to the case~\(L^s=1^0\) (this also corresponds to the behavior of a single, uncoupled unit).
The observed transition of regimes corresponds to the one thoroughly studied before~\cite{karnatak2014} and can be explained by the two networks behaving like a single network.
However, this does not apply to the dynamics within the atypical region described in the following section.

\begin{figure}
\includegraphics[width=\linewidth]{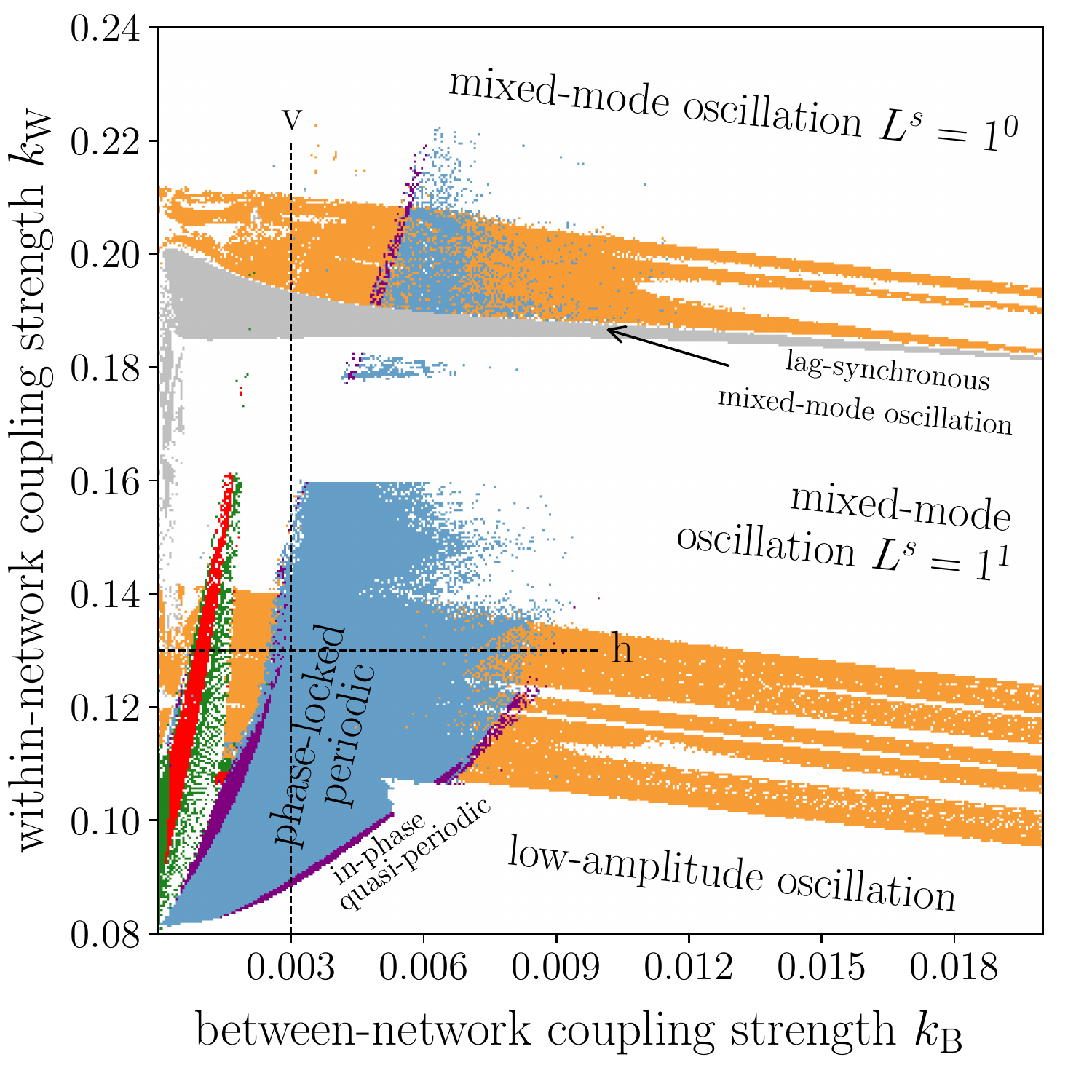}
	\caption{Enlargement of the parameter range encircled by the dashed curve in Fig.~\ref{fig:blob1}, with additional specification of the atypical region:
	Red denotes anti-phase-locked periodic oscillations;
	green denotes anti-phase quasi-periodic oscillations;
	gray denotes lag-synchronous mixed-mode oscillations;
	blue denotes phase-locked periodic oscillations;
	purple denotes in-phase quasi-periodic oscillations;
	orange denotes chaotic dynamics.
	The remaining dynamics are declared by text.
	The horizontal cut (h) is studied in detail in Sec.~\ref{sec:HorizCut} and can be seen in Fig.~\ref{fig:HorizontalCut}.
	The vertical cut (v) is studied in detail in Sec.~\ref{sec:VertiCut} and can be seen in Fig.~\ref{fig:VerticalCut}.	
	Unlike Fig.~\ref{fig:blob1}, each simulation uses the same initial condition, in order to ensure that the system remains approximately on the same manifold and the regions of different dynamics are better captured.}\label{fig:blob2}
\end{figure}

\section{Dynamics in the atypical region}\label{sec:Cuts}
\subsection{Overview of dynamics}\label{sec:Cuts:Overview}
For a weak coupling between sub-networks (\(k_\text{B} \in \kle{0.0,0.02}\)) and for a range of within-network couplings (\(k_\text{W} \in \kle{0.08,0.22}\)) (the black region in Fig.~\ref{fig:blob1}, zoomed in Fig.~\ref{fig:blob2}), we observe the dynamics of sub-networks to strongly differ from the aforementioned symmetric structure, where the sub-networks behave each as independent, but separate networks. 
Note that a rigorous outlining of the atypical region is hindered by the presence of multistability, discussed in Sec.~\ref{sec:Multistability}.

More specifically, we observe the following dynamics in the atypical region (also see Figs. \ref{fig:plots} and~\ref{fig:blob2}):
\begin{description}
	\item[in-phase quasi-periodic]
		Both sub-networks exhibit low-amplitude oscillations whose amplitudes differ over time and between the sub-networks.
		There is a phase offset between the two sub-networks with a varying phase difference.
	\item[phase-locked periodic]
		Both sub-networks exhibit low-amplitude oscillations whose amplitudes are constant over time but differ between the sub-networks.
		There is a constant phase offset between the two sub-networks.
	\item[anti-phase-locked periodic]
		The two sub-networks exhibit low-amplitude oscillations with the same amplitude offset by half a period.
	\item[anti-phase quasi-periodic]
		Both sub-networks exhibit low-amplitude oscillations whose amplitudes differ over time and between the sub-networks.
		The phase offset varies over time, but never becomes zero.
	\item [lag-synchronous mixed-mode oscillations]
		Each sub-network exhibits the same mixed-mode oscillation, but there is a constant phase offset between the sub-networks.
\end{description}
The phase-locked periodic dynamics is the only one where one sub-network can consistently \emph{dominate} the other, in the sense that it has a persistently larger oscillation amplitude than the other network.
As expected from the symmetry of the scenario, it entirely depends on initial conditions which sub-network is dominant, i.e., the system is bistable.
However, we found no means to predict the dominant sub-network from the initial conditions.
We also observed cases where the dominance of the sub-networks switches, but only in regimes of chaotic dynamics.

\subsection{Parameter cuts through the atypical region}\label{sec:cuts}
In order to thoroughly study the underlying dynamics in the atypical region and the bifurcations governing the transitions between these dynamics, and to showcase the sequences of dynamical regimes near or within this region, we perform cuts through the two-dimensional parameter region, varying only one of the coupling strengths.
We verify the bifurcations occurring at each transition point by performing a numerical continuation of the various stable and unstable fixed points and limit cycles~\cite{doedel1991,ermentrout2002animating}.

The nature of the various dynamical regimes observed is characterized using the three largest Lyapunov exponents \(\lambda_1\), \(\lambda_2\), and \(\lambda_3\).
We classify the dynamics as \emph{multistable} for a given parameter setting if we do not observe the same dynamical behavior over \(25\)~random initial conditions.
Note from Fig.~\ref{fig:plots} that for the parameter regime under consideration, the dynamics of the two sub-networks may not be identical.
For any such dynamics, there is an inherent ``trivial'' bistability arising from the identical nature of the sub-networks.
In the following sections of the paper, we exclude this bistability when speaking of \emph{multistability}.

\begin{figure}
	\includegraphics[width=\linewidth]{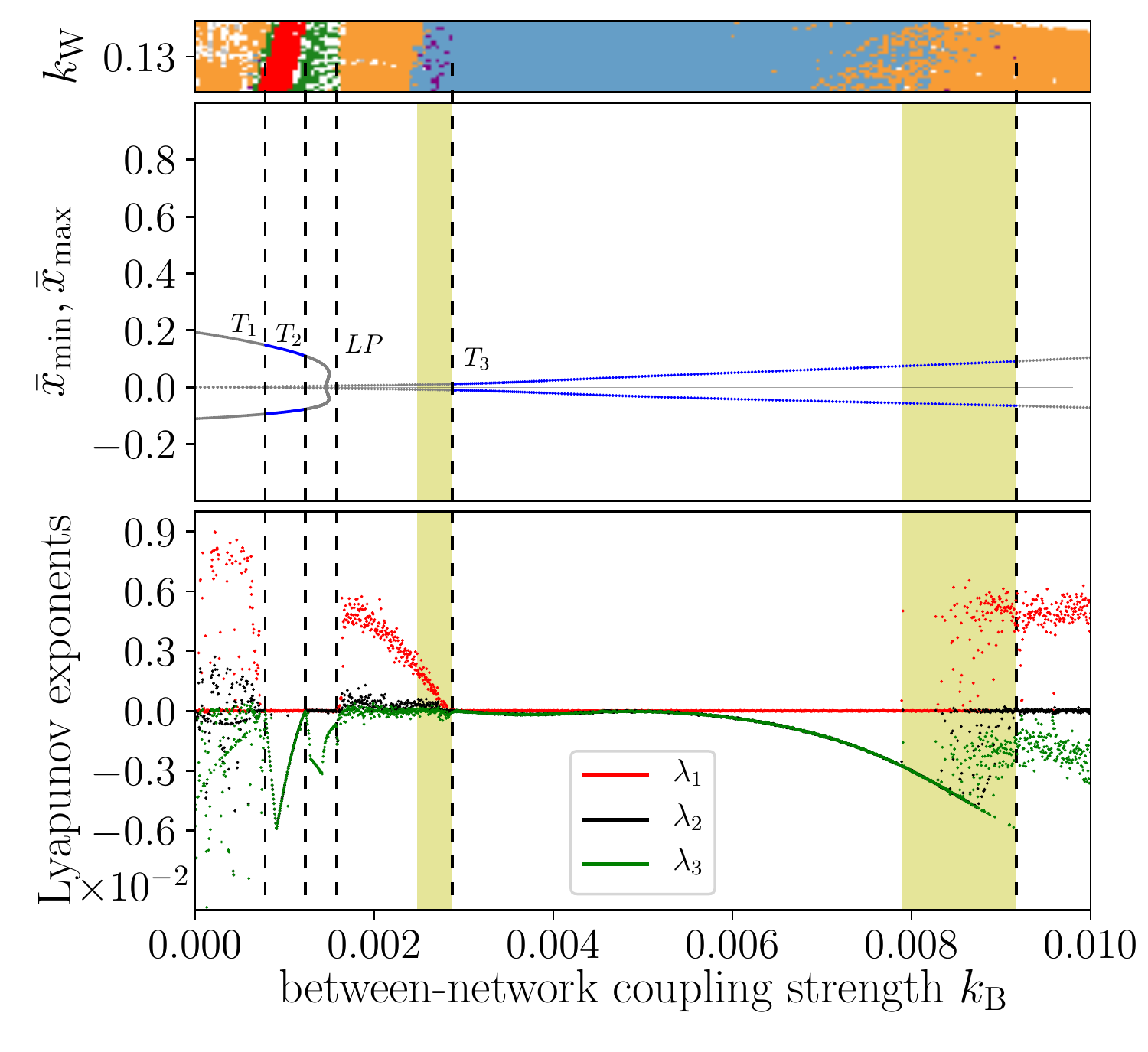}
	\caption{Parameter regimes for \(k_\text{W}=0.13\) (horizontal cut (h) in Fig.~\ref{fig:blob2}).
	Top: Excerpt of Fig.~\ref{fig:blob2} showing the neighborhood of the cut.
	Middle: Minimal and maximal values of the average observable \(\bar{x}\).
	The blue dots represent stable solutions of the numerical continuation, and the gray unstable.
	Bottom: The three largest Lyapunov exponents of the system.
	The bifurcations are marked by the dashed vertical lines, including torus bifurcations ($T_1$, $T_2$, $T_3$)	and a limit point~\textit{(LP).}
	Regimes of multistability are shaded yellow.
	The top and middle panel use unique initial conditions; the bottom panel uses differing ones.
	All data is based on \(10^5\) time units.
	}
	\label{fig:HorizontalCut}
\end{figure}

\subsubsection{Varying \(k_\text{B}\)}\label{sec:HorizCut}

Fixing the within-network coupling strength at \(k_\text{W}=0.13\), we vary the between-network coupling strength \(k_\text{B}\) from \(0\) to \(0.01\).
This transversal cut covers a wide variety of dynamics (see Fig.~\ref{fig:HorizontalCut}).

For \(k_\text{B}=0\), the uncoupled sub-networks perform independent mixed-mode oscillations.
As the coupling is slightly increased, the sub-networks start interacting and exhibit a variety of chaotic dynamics.
For \(0.000780 \lesssim k_\text{B} \lesssim 0.001229\), the dynamics stabilizes into a limit cycle (\(\lambda_1=0\); \(\lambda_2=\lambda_3<0\)), giving rise to anti-phase-locked periodic oscillations, via a reverse torus bifurcation~$T_1$.
This stable limit cycle loses its stability at \(k_\text{B} \approx 0.001229\) via a torus bifurcation~$T_2$, giving rise to a stable torus on which the system executes anti-phase quasi-periodic oscillations (\(\lambda_1=\lambda_2=0\); \(\lambda_3<0\)).

At \(k_\text{B} \approx 0.00158\), the unstable limit cycle around which the corresponding torus is formed, disappears together with the torus in a limit-point bifurcation \textit{(LP).}
Therefore, the dynamics becomes chaotic (\(\lambda_1>0\)) with occasional high-amplitude oscillations.
As we move closer to \(k_\text{B} \approx 0.002872\), the largest Lyapunov exponent \(\lambda_1\) decreases and the high-amplitude oscillations become progressively less frequent.
On increasing \(k_\text{b}\) further, we reach a small range of parameter \(0.00248 \lesssim k_\text{B} \lesssim 0.002872\) where multistability is observed in our numerical simulations.
Depending on the initial conditions, the trajectories may converge to either in-phase quasi-periodic oscillations, phase-locked periodic oscillations or chaotic oscillations (see the top panel of Fig.~\ref{fig:HorizontalCut}).
As we reach \(k_\text{B} \approx 0.002872\), the quasi-periodic motion ceases via a reverse torus bifurcation $T_3$ and all trajectories converge to phase-locked periodic dynamics, i.e., a limit cycle (\(\lambda_1=0; \lambda_2=\lambda_3<0\)).

This limit cycle remains stable until \(k_\text{B} \approx 0.009169\) after which the trajectories revert back to exhibiting chaotic motion involving occasional high-amplitude oscillations (\(\lambda_1>0; \lambda_2=0; \lambda_3<0\)).
Note that the attractor corresponding to chaotic dynamics appears already at \(k_\text{B} \approx 0.0079\) and continues beyond \(k_\text{B} = 0.01\).
The system is therefore multistable in the range of \(0.0079 \lesssim k_\text{B} \lesssim 0.009169\), where a trajectory can converge either to the small-amplitude limit cycle, i.e., phase-locked periodic oscillations, or to the chaotic attractor.

\begin{figure}
	\includegraphics[width=\linewidth]{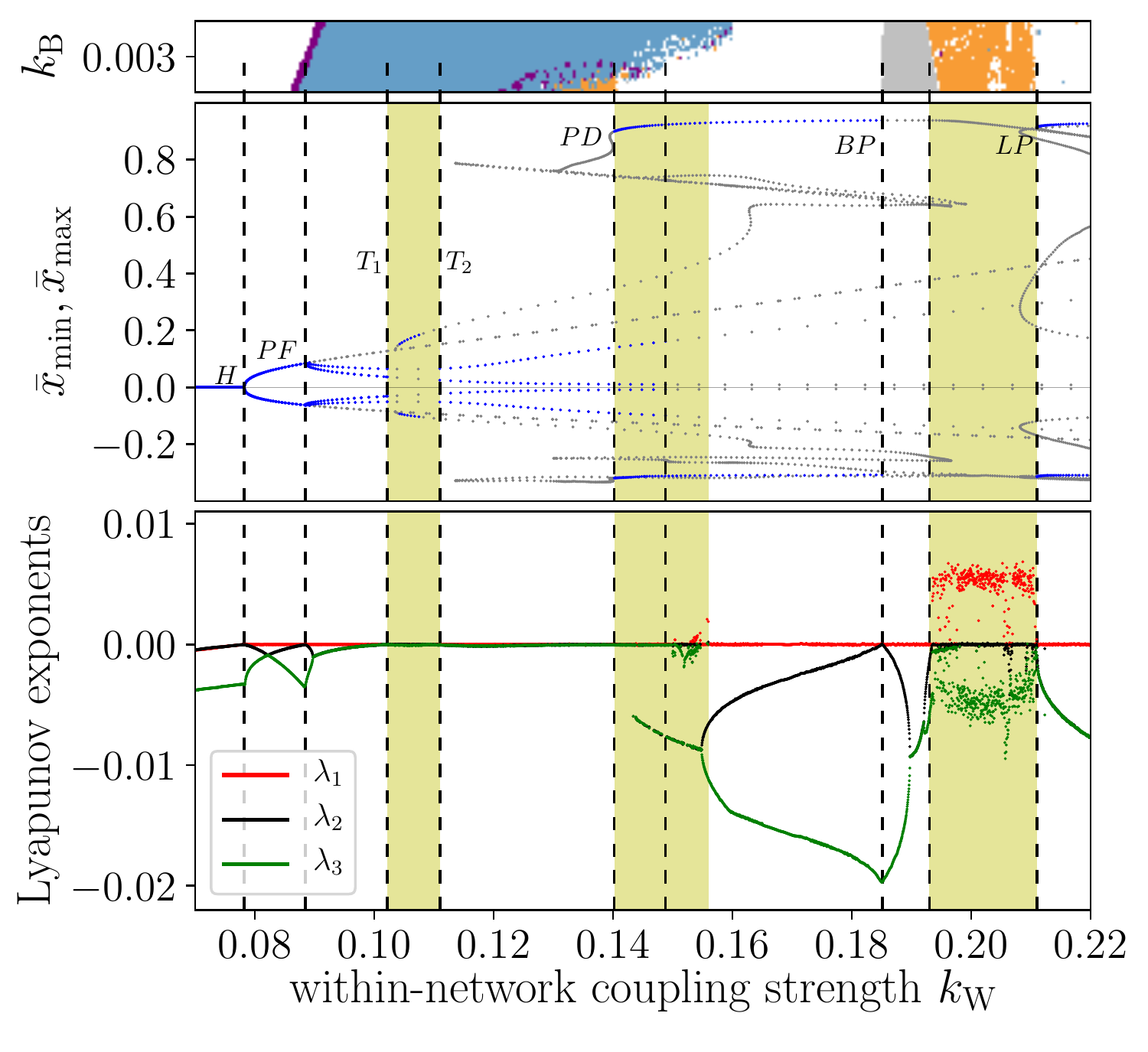}
	\caption{Parameter regimes for \(k_\text{B}=0.003\) (vertical cut (v) in Fig.~\ref{fig:blob2}).
	Top: Rotated excerpt of Fig.~\ref{fig:blob2} showing the neighborhood of the cut.
	Middle: Minimal and maximal values of the average observable \(\bar{x}\).
	The blue dots represent stable solutions of the numerical continuation, and the gray unstable.
	Bottom: The three largest Lyapunov exponents of the system.
	The bifurcations are marked by the dashed vertical lines, including a Hopf bifurcations~\textit{(H),} a pitchfork bifurcation~\textit{(PF),} torus bifurcations ($T_1, T_2$),	a period doubling~\textit{(PD),} a branch point~\textit{(BP)} and a limit point~\textit{(LP).}
	Regimes of multistability are shaded yellow.
	The top and middle panel use unique initial conditions; the bottom panel uses differing ones.
	All data is based on \(10^5\) time units.}
	\label{fig:VerticalCut}
\end{figure}

\subsubsection{Varying \(k_\text{W}\)}\label{sec:VertiCut}
We observe an even more intricate sequence of bifurcations when the within-network coupling strength \(k_\text{W}\) is increased, whilst the between-network coupling strength \(k_\text{B}\) is fixed at \(0.003\) (see Fig.~\ref{fig:VerticalCut}).

When \(k_\text{W} \rightarrow 0\), the units of the two sub-networks are connected effectively only by the between-network couplings.
The origin of state space is the only stable fixed point in the system (\(\lambda_1<0\)) and all trajectories converge to this global attractor.
It loses its stability in a supercritical Hopf bifurcation~\textit{(H)} at \(k_\text{W} \approx 0.07823\), beyond which all trajectories converge to the newly formed limit cycle (\(\lambda_1=0; \lambda_2,\lambda_3<0\)), where both sub-networks exhibit synchronous low-amplitude oscillations.
At \(k_\text{W} = 0.08849\), this limit cycle becomes unstable through a pitchfork bifurcation~\textit{(PF),} giving rise to phase-locked periodic oscillations which are no longer synchronized.

Numerical continuation of the limit cycles shows that the stability of the two limit cycles is lost via a torus bifurcation $T_1$ at \(k_\text{W} \approx 0.1022\) and regained only at \(k_\text{W} \approx 0.111\), via a reverse torus bifurcation $T_2$. 
Therefore, in between these bifurcation points, the system is expected to exhibit quasi-periodicity.
While the Lyapunov exponents corroborate the results of the continuation methods (with \(\lambda_1=\lambda_2=0; \lambda_3<0\) for this parameter range), in the numerical simulations using a single initial condition, 
we did not find those regions of quasi-periodicity.
Note that, in addition to the tori, a stable limit cycle also appears for the parameter interval \(0.104 \lesssim k_\text{W} \lesssim 0.1075\).
This results in the system being multistable in the aforementioned intervals of parameters.

The limit cycles which emerge after $T_2$ remain stable until \(k_\text{W} \approx 0.1488\).
Within this parameter window, another stable high-amplitude limit cycle corresponding to mixed-mode oscillations emerges at \(k_\text{W} \approx 0.1402\), via a period doubling~\textit{(PD).}
Note that this high-amplitude limit cycle lies on the manifold corresponding to complete synchrony of the two sub-networks.
The stability of this limit cycle extends until \(k_\text{W} \approx 0.1851\), vanishing at a branch-point~\textit{(BP).}
This implies that the system shows a second region of multistability when \(0.1402 \lesssim k_\text{W} \lesssim 0.156\).
The existence of both limit cycles is also indicated by the Lyapunov exponents.
Throughout this parameter range, the largest Lyapunov exponent \(\lambda_1\) is zero, and all the other exponents are negative.
The region of multistability is also indicated by the fact that the second and third largest Lyapunov exponents assume different values depending on the initial conditions.

Beyond \(k_\text{W} \approx 0.1851\) the system exhibits lag-synchronous mixed-mode oscillations (\(\lambda_1=0; \lambda_2,\lambda_3<0\)) of type \(L^s=1^1\), which lose stability and give rise to a chaotic state ($\lambda_1>0$) within the interval \(0.193 \lesssim k_\text{W} \lesssim 0.211\).
Small regions of multistability exist in this area.
The system regains stability via a limit-point bifurcation~\textit{(LP),} entering a synchronous mixed-mode oscillations of type \(L^s=1^0\), i.e., pure high-amplitude oscillations.
This is the stable dynamics observed beyond $k_\text{W} > 0.211$.

\begin{figure}
\includegraphics[width=\linewidth]{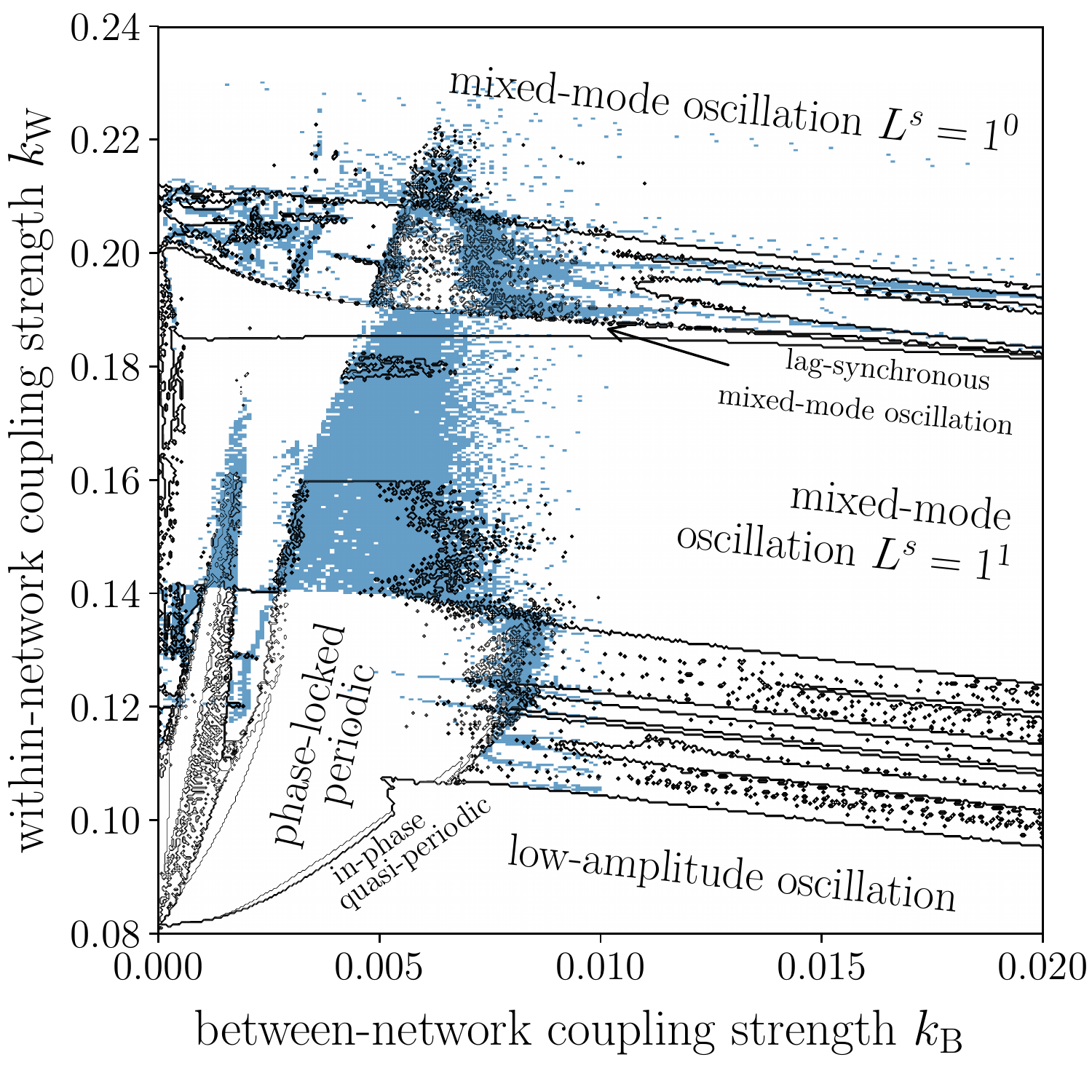}
	\caption{Regions of multistability (blue) in the system as estimated from \(25\)~initial conditions.
	Contours of the parameter regimes (black) as in Fig.~\ref{fig:blob2} for orientation.
	Multistability exists mainly surrounding the atypical region, i.e., most dynamical regimes of a single network exist alongside the novel dynamics emerging from the novel dynamics of system Eq.~\ref{Eq:FHN}.
	Random initial conditions are taken at each iteration, in order to uncover multistability.
	}
	\label{Multistability}
\end{figure}

\subsection{Multistability}\label{sec:Multistability}
As can be seen already in Figs.~\ref{fig:HorizontalCut} and~\ref{fig:VerticalCut}, the parameter values at which bifurcations occur differ slightly between the results obtained from simulations and those from employing continuation methods.
This is due to the existence of multistability in the system.

In Fig.~\ref{Multistability}, we show the regions in parameter space where multistability occurs.
We observe that the borders of transition to multistability partially match the borders of the atypical region.
The exact transitions and stability of the system are not specified, since the goal is to portray the connection of the dynamics in the atypical region to the emergence of multistability.

\begin{figure}
\includegraphics[width=\linewidth]{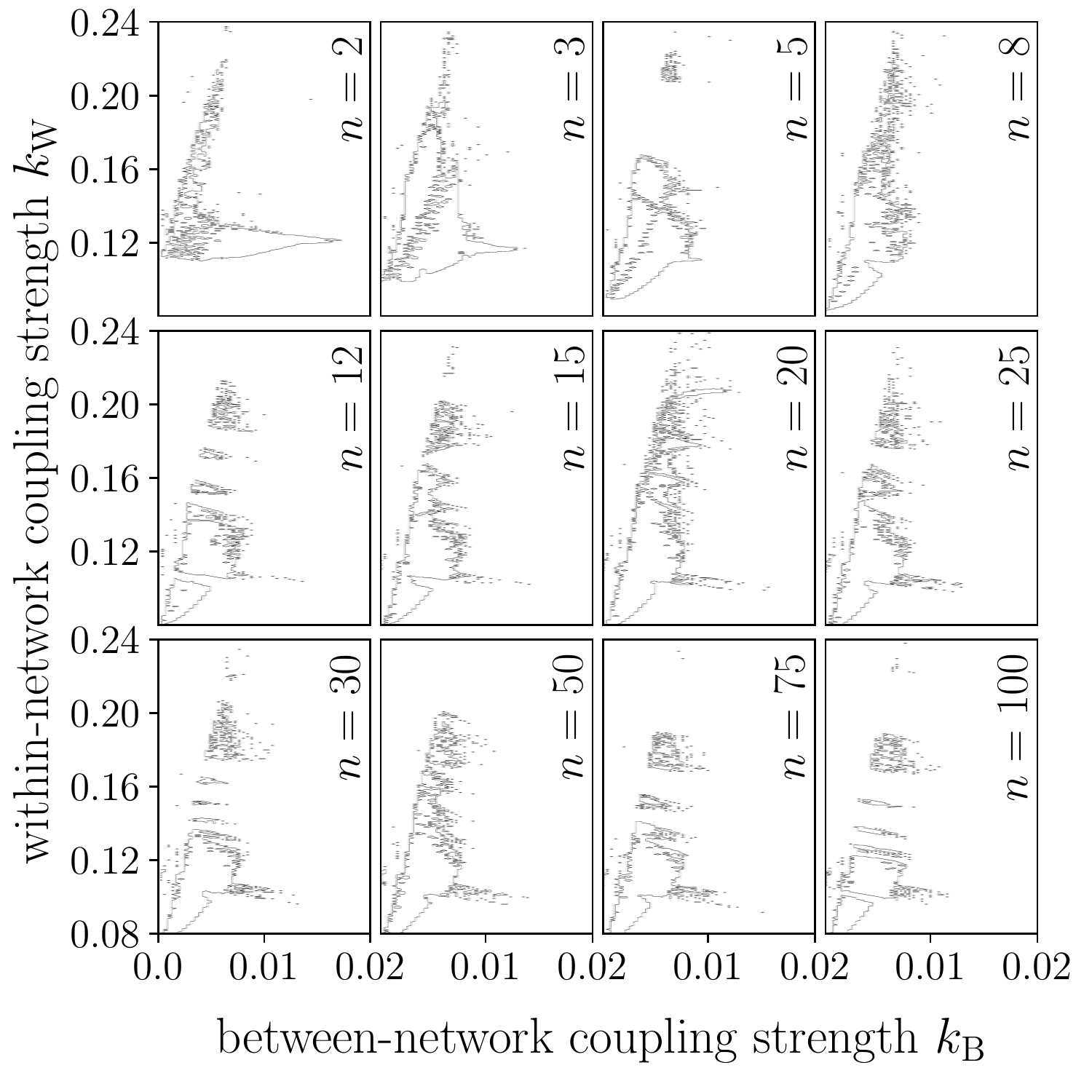}	
\caption{
		Boundaries of the in-phase quasi-periodic and phase-locked periodic regimes (purple and blue areas in Fig.~\ref{fig:blob2}) for different numbers of units per sub-network~(\(n\)).
		The atypical regions is present in the normalized parameter space ($k_{\mathrm{B}},k_{\mathrm{W}}$) in all cases, but can exhibit slightly different shapes.
		}
	\label{Sizeplay}
\end{figure}

\subsection{Dependence on sub-network size}\label{sec:Sizeplay}
To ensure the above results are not singular to the number of units in either sub-network, we investigated analogous systems with \(n=2\) to \(n=100\) units in each sub-network.
Note that this comparison is facilitated through the normalizations with the sub-networks' size~\(n\) in Eqs.~\ref{Eq:b},~\ref{Eq:internal_coupling}, and~\ref{Eq:external_coupling}.
In Fig.~\ref{Sizeplay}, we show how the sub-network size affects the parameter regimes exhibiting in-phase quasi-periodic and phase-locked periodic regimes.
We chose these regimes for a comparison as they are largest and have comparably sharp borders.
There are of course several other dynamics involved, but our aim is here to demonstrate the existence of the described phenomena for different sub-network sizes and not to detail the dynamics or bifurcations.
Since both parameter regimes display a roughly uniform structure across all investigated sub-network sizes, the respective dynamics can thus be observed for all these cases, concluding that the phenomena exist independently from the system's size.

\section{From networks of networks of excitable units to coupled complex oscillators}
\begin{figure}
\includegraphics[width=.6\linewidth]{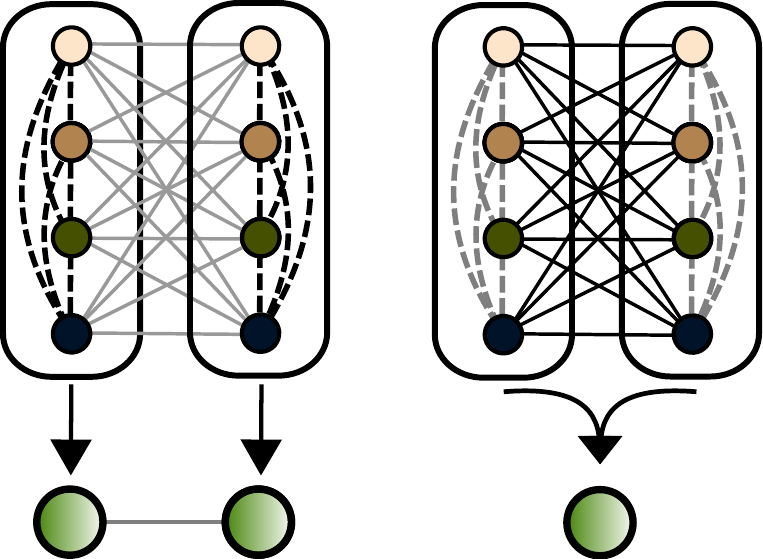}
	\caption{	
	Sketch of the effective coupling structure for two \(n=4\) sized sub-networks. 
	The solid lines indicate the between-network coupling with strength \(k_{\mathrm{B}}\), dashed lines indicate within-network coupling with strength \(k_{\mathrm{W}}\).
	The black and gray coloring indicate strong or weak coupling respectively.
	Left: For a small~\(k_{\mathrm{B}}\), the all-to-all topology becomes negligible, particularly between the sub-networks, which results in two irreducible, interacting higher-dimensional oscillators.
	An example for such a case is a phase-locked periodic oscillation as obtained for $(k_{\mathrm{W}},k_{\mathrm{B}})=(0.1,0.0003)$ (see Fig.~\ref{fig:plots}, sixth column).
	Right: For sufficiently large values of~\(k_{\mathrm{B}}\), the all-to-all coupling is effective, making the network of sub-networks reducible to a single network/oscillator.	
	All dynamics outside the borders of Fig.~\ref{fig:blob2}, i.e., outside the parameter range encircled by the dashed lines in Fig.~\ref{fig:blob1} are examples of such.
	Each unit color denotes a different intrinsic parameter \(b_i\) of a unit.
	The sub-networks are structurally identical.
	}
\label{Fig:Last}
\end{figure}
In order to facilitate an understanding of the observed phenomena, we now discuss how varying the within-network and between-network
coupling strength affects the effective structure of our network.
If the between-network coupling~\(k_{\mathrm{B}}\) is small (but not zero), the sub-networks interact only weakly.
If additionally \(k_{\mathrm{W}}\)~is sufficiently large, i.e., inside the atypical region, each sub-network can be viewed as a single, independent higher-dimensional oscillator.
Phase-locked periodic, in-phase quasi-periodic, phase-locked periodic, anti-phase-locked periodic and lag-synchronous mixed-mode oscillations belong to this category.
Our network of networks can thus be considered as a system of two coupled identical oscillators, each with a rich dynamical behavior, and the coupling structure leads to the observed dynamics in the atypical region in parameter space (see Fig.~\ref{Fig:Last}, left).
This explains why the observed dynamical behaviors are richer than those for a single network~\cite{ansmann2013, karnatak2014}.

On the other hand, if the between-network coupling~\(k_{\mathrm{B}}\) is sufficiently large, the all-to-all coupling dominates the dynamics and the two sub-networks merge.
All dynamics where one cannot distinguish the state variables $\bar{x}_1, \bar{x}_2$ belong to this category.
This includes all dynamics with $k_{\mathrm{B}}>0.02$.
Low-amplitude oscillations and mixed-mode oscillations are examples of such a dynamics.
Hence, the above point of view is no longer valid, and our network of networks (with \(n\)~units each) behaves like a single network of \(2n\)~units, with a modified within-network coupling strength (see Fig.~\ref{Fig:Last}, right).

\section{Conclusion}
\label{sec:conclusion}
We investigated the dynamics of a network of all-to-all coupled, identical sub-networks consisting of diffusively coupled, non-identical excitable units.
Such a configuration naturally leads to the question whether an interdependent network is not just one bigger network whose (global) dynamics can trivially be deduced from the dynamics of its components.
We identified a region in parameter space, in which the interplay of within- and between-network couplings allows for a richer dynamical behavior than can be observed for a single sub-network.
The existence of this region turned out to be robust for a wide range of sub-network sizes.
It is characterized by a within-network coupling that is large in relation to the between-network coupling.

In some aspects of the dynamics, we find a typical behavior of two coupled oscillators such as emergence of quasi-periodicity and phase-locked motion on the torus.
In other aspects of the dynamics, such as mixed-mode oscillations, the system behaves just like a bigger network.
Both aspects overlap in parameter space leading to multistability and exhibiting the two faces of dynamics as (i)~just a bigger network and as (ii)~two coupled ``oscillators'', where each such oscillator is a sub-network.

The emergence of novel dynamics in this region indicates that approximations of a network of networks by a single (but bigger) network might be not accurate enough to properly understand the complexity of dynamics on networks of networks.
Instead we find two distinct behaviors~--~one attributed to the coupling of two oscillators (sub-networks) and another representing the bigger network~--~which coexist in a region of multistability.
Our findings thus support recent perspectives on the irreducibility when representing interdependent systems as networks of networks, such as in spreading processes~\cite{dedomenico2016b},
ecological networks~\cite{pilosof2017},
biochemical networks~\cite{klosik2017},
transportation networks~\cite{kleineberg2016,osat2017},
or the brain~\cite{dedomenico2016} and underline the necessity for network of networks analyses of the underlying architecture.

\section*{Acknowledgments}
The authors would like to thank H.~Dickten, C.~Geier, J.~Heysel, T.~Rings, and K.~Simon for interesting discussions.
This work was partially supported by the Volkswagen Foundation (Grant~Nos. 88459 and 88463).

\end{document}